\documentclass[showpacs,twocolumn,nobalancelastpage, superscriptaddress, 10pt,aps, prstab]{revtex4-1}

\usepackage{amssymb,amsfonts,amsmath}
\usepackage{graphicx}
\usepackage{color}
\usepackage{epsfig}
\usepackage{dsfont}
\usepackage{braket}

\begin{document}

% Title Page
\title{Magnetic adatoms as memory bits: A quantum master equation analysis}

\author{Christian Karlewski}
\email{christian.karlewski@kit.edu}
\affiliation{Institut f\"ur Theoretische Festk\"orperphysik, Karlsruhe Institute of Technology, D-76131 Karlsruhe, Germany}
\affiliation{Institute of Nanotechnology, Karlsruhe Institute of Technology, D-76344 Eggenstein-Leopoldshafen, Germany}
\author{Michael Marthaler}
\affiliation{Institut f\"ur Theoretische Festk\"orperphysik, Karlsruhe Institute of Technology, D-76131 Karlsruhe, Germany}
\author{Tobias M\"arkl}
\affiliation{Physikalisches Institut, Karlsruhe Institute of Technology, D-76131 Karlsruhe, Germany}
\author{Timofey Balashov}
\affiliation{Physikalisches Institut, Karlsruhe Institute of Technology, D-76131 Karlsruhe, Germany}
\author{Wulf Wulfhekel}
\affiliation{Institute of Nanotechnology, Karlsruhe Institute of Technology, D-76344 Eggenstein-Leopoldshafen, Germany}
\affiliation{Physikalisches Institut, Karlsruhe Institute of Technology, D-76131 Karlsruhe, Germany}
\author{Gerd Sch\"on}
\affiliation{Institut f\"ur Theoretische Festk\"orperphysik, Karlsruhe Institute of Technology, D-76131 Karlsruhe, Germany}
\affiliation{Institute of Nanotechnology, Karlsruhe Institute of Technology, D-76344 Eggenstein-Leopoldshafen, Germany}


\date{\today}
\begin{abstract}
Due to underlying symmetries the ground states of magnetic adatoms may be highly stable, which opens perspectives for application as single-atom memory. 
A specific example is a single holmium atom (with $J=8$) on a platinum (111) surface for which exceptionally long lifetimes were observed in recent scanning tunneling microscopy studies.
For control and read-out the atom must be coupled to electronic contacts.
Hence the spin dynamics of the system is governed by a quantum master equation.
Our analysis shows that in general it cannot be reduced to a classical master equation 
in the basis of the unperturbed crystal-field Hamiltonian. Rather, depending on parameters and control fields, ``environment induced superselection" principles choose the appropriate set of basis states, which in turn determines the specific relaxation channels and lifetimes. Our simulations suggest that in ideal situations the lifetimes should be even longer than observed in the experiment. We, therefore, investigate the influence of various perturbations. We also study the initialization process of the state of the Ho atom by applied voltage pulses and conclude that fast, high fidelity preparation, on a $100\,\text{ns}$ timescale, should be possible.
\end{abstract}

\maketitle

\section{Introduction}
\begin{figure}[hbt]
 \includegraphics[origin=c,width=0.33\textwidth]{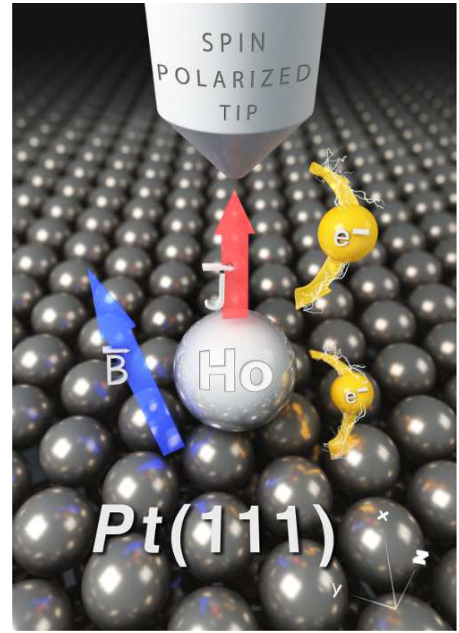}
 \caption{Setting of the experiment in Ref.\ \cite{nature12759}: The magnetic state of an holmium atom on a Pt(111) surface is studied by a scanning tunneling microscope with a spin-polarized tip.}
  \label{pSetting}
\end{figure}

An ultimate boundary of miniaturization of information technology is reached when single atoms are used as memory bits. In this respect the experiment of Miyamachi {\sl et al.}~\cite{nature12759} with a single Ho atom positioned on a Pt(111) surface (as depicted in Fig.~\ref{pSetting}) represent an important milestone, promising  lifetimes of several minutes for the two degenerate magnetic ground states. 
For the crystal-field Hamiltonian with parameters as determined in Ref. \cite{nature12759}, the two ground states of the Ho atom have $\langle J_z \rangle \approx\pm8$ pointing into or out of the metal surface. The  long lifetimes result from a combination of symmetries and specific properties of the system \cite{nature12759}. 
They are many orders of magnitude longer than those measured for single Co/Fe atoms on Pt(111) \cite{PhysRevLett.106.037205, PhysRevLett.102.257203} or other magnetic adatoms on a metallic surface \cite{Science-2010-Loth-1628-30}. 
 A related giant magnetic anisotropy of single adatoms had been observed earlier for Co atoms on Pt(111) \cite{Science-2003-Gambardella-1130-3}.
We should mention that in another recent work with Ho on Pt(111), 
Donati {\sl et al.} \cite{PhysRevLett.113.237201} arrived at different crystal-field parameters, leading to ground states with $\langle J_z \rangle \approx\pm6$ and much shorter lifetimes.

In this paper we develop the theory how to treat magnetic adatoms on metallic surfaces which -- for read-out and control -- are coupled to electronic reservoirs formed by the metal substrate and a spin-polarized STM tip. The description of this dissipative system is based on a quantum master equation for the reduced density matrix for the angular momentum states, which accounts for the influence of the electronic contacts and the applied voltage.
For low voltages the description reduces to a rate equation in the eigenbasis of the unperturbed crystal-field Hamiltonian, which are specific superpositions of angular momentum states. However, for larger voltages the coherence of these superposition states is destroyed, and the relevant states are the steady-state eigenstates of the full dissipative quantum master equation. This is an example for what has been called ``environment-induced superselection" (or short ``einselection") principle \cite{ProgZurek-281-312,RevModPhys.75.715}. In the considered system the choice of the appropriate basis depends strongly on the voltage. It strongly influences the relaxation channels and times. Switching the voltages
also allows initializing the systems in a specific ground state, which is needed when using the system as memory.

As a specific example we analyze Ho on Pt(111) with
crystal-field Hamiltonian  as determined in Ref.~\cite{nature12759}, but we also compare with the situation obtained for the parameters of Ref.~\cite{PhysRevLett.113.237201}. 
The quantitative comparison with the experiments requires assumptions about a number of 
further parameters, which we explore in the latter part of this paper in some detail. For the comparison we also present  experimental data which resolve, beyond what has been reported in Ref.~\cite{nature12759}, the lifetimes for each one of the low-energy states.

\allowdisplaybreaks
\section{The model}
\subsection{The crystal-field Hamiltonian and eigenstates}
We investigate the dynamics of a magnetic adatom placed on a high-symmetry position on a metal surface in a situation where the symmetry stabilizes the degenerate ground states against the dominant perturbations. Such a situation is realized for Ho on Pt(111), which we consider as a specific example, but our analysis and conclusions can easily be generalized to other magnetic adatoms. Ho adatoms on a Pt(111) surface have been investigated experimentally in Refs. \cite{nature12759, 1-s2.0-S0039602814002180-main,PhysRevLett.113.237201}. The Ho atom has strong spin-orbit coupling, therefore the total angular momentum is a good quantum number with $J= 8$, leading to 17 states in the multiplet to be studied. A single adatom on a crystal structure with trigonal symmetry  can be described by a crystal-field Hamiltonian adjusted to the symmetries \cite{0034-4885_16_1_304,99324},
\begin{align}
	\label{eHCF}
	H_{CF}=& \sum_{\substack{n=2,4,6\\m=0,3,6\\m\leq n}}B_n^mO_n^m\\
	=& \  B_2^0\cdot 3\, J_z^2 +B_4^0\cdot35\,J_z^4\notag\\
	&+ B_4^3\cdot\frac{1}{4}\left[J_z,(J_+^3+J_-^3)\right]_+\ldots \, .\notag
\end{align}
Here $O_n^m$ are the Stevens operators expressed in powers of $J_z$, $J_+$ and $J_-$ angular momentum operators of the Ho atom \cite{Wybourne1965}. Due to the trigonal symmetry the operators $J_+$ and $J_-$ appear in powers of multiples of three. 
Above, the first three terms are shown explicitly (with $[\cdot,\cdot]_+$ denoting the anti-commutator), further ones are listed in the Appendix. 
The coefficients $B_n^m$ are crystal-field parameters, which in Ref.~\cite{nature12759} were determined from {\sl ab-initio} calculations and compared with experiment. 
For Ho and many other magnetic adatom systems the leading term is $\propto J_z^2$ with a negative coefficient $B_2^0 <0$.
The difference in the conclusions reached in 
Refs.~\cite{nature12759} and \cite{PhysRevLett.113.237201} arise from different values of the parameter $B_4^0$. We first proceed using the values of Ref.~\cite{nature12759} but we will comment of the situation of Ref. \cite{PhysRevLett.113.237201} in section \ref{cDonati}.

\begin{figure}[hbt]
 \includegraphics[origin=c,width=0.48\textwidth]{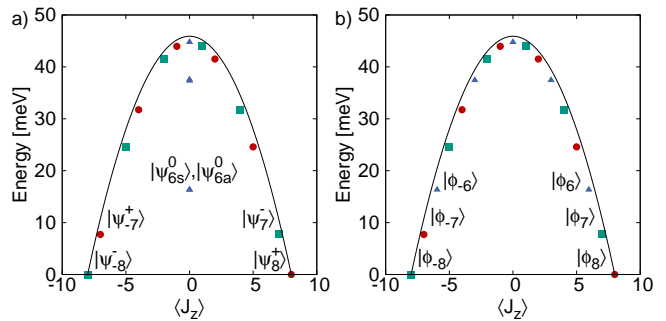}
 \caption{a) Energies of the eigenstates of the Hamiltonian $H_{CF}$
plotted versus the expectation value  $\langle J_z\rangle$. b) Energy expectation values of the steady states $\ket{\phi_m}$ of the quantum master equations for strong dissipation due to an applied voltage of $U=7.3\,\text{mV}$. The labels refer to the dominantly contributing $J_z$-eigenstates.}
  \label{pLevels}
\end{figure}

The eigenstates of the crystal-field Hamiltonian $H_{CF}$ can be divided into three families of states, $\ket{\psi^+_m}$, $\ket{\psi^-_m}$, and $\ket{\psi^0_m}$. Each one of these states is a superposition of different $J_z$ eigenstates with magnetic quantum numbers differing by multiples of 3.
The lower index $m$ of each state, with $-8 \le m \le 8$, denotes the dominantly contributing $J_z$ eigenstate. 
In Fig.~\ref{pLevels} we plot the eigenenergies versus the expectation value
$\langle J_z \rangle$. They are marked by circles, squares and triangles for the three families $+,-,0$. 
Since the dominant contribution to the eigenenergies arises from the first term $\propto -J_z^2$ the energies lie approximately on an inverted parabola. 
Note that by plotting the energy versus $\langle J_z \rangle$ we can also represent arbitrary superposition states of the aforementioned basis states, which is useful for the following discussions.

The two degenerate ground states of the system $\ket{\psi_{8}^+}$ and $\ket{\psi_{-8}^-}$, 
 with differing angular moment pointing into or out of the surface, belong to two different 
families ($+$ and $-$). The same applies for the first excited states $\ket{\psi_{7}^+}$ and 
$\ket{\psi_{-7}^-}$,  as well as for various higher ones.  
These  states are superposition of non-degenerate  $J_z$-eigenstates. In contrast, the states $\ket{\psi^0_{6s}}$ and $\ket{\psi^0_{6a}}$, belonging to the $0-$family, are superpositions made up of degenerate  $J_z$-eigenstates, coupled by the operators 
$O_4^3$, $O_6^3$ and $O_6^6$. They naturally split into symmetric and antisymmetric (s,a) combinations, both with vanishing $\langle J_z \rangle$, and are marked accordingly in Fig.~\ref{pLevels}~a). The same applies for $\ket{\psi^0_{3s}}$ and $\ket{\psi^0_{3a}}$.

\subsection{Spin-dependent electron scattering}
For control and read-out of the magnetic state the Ho adatom is coupled to a spin-polarized STM tip. The electronic reservoirs influence the dynamics of the system by the scattering of electrons in the electrodes from the  atom and by the tunneling of electrons between tip and bulk substrate via the atom (see Fig.~\ref{pSetting}). Accordingly the total Hamiltonian consists of three parts,  $H=H_S+H_{Res}+H_C$, describing the system, the electron reservoirs in the bulk substrate and tip of the STM, and the coupling terms,
\begin{align}
 H_S=&\, H_{CF}+g_J\mu_B\vec{B}\cdot\vec{J}, \label{system}\\
 H_{Res}=&\sum_{\substack{\alpha=T,B\\\sigma=\uparrow,\downarrow; \, k}}
\left(\varepsilon^\alpha_{k \sigma}+eU^\alpha\right)\, c_{k\sigma}^{\alpha\dagger} c_{k\sigma}^\alpha,\\
 H_C= &\sum_{\substack{\alpha,\alpha'=T,B\\kk'}}  t_{kk'}^{\alpha\alpha'}  \left[J_+c_{k\downarrow}^{\alpha\dagger}c_{k'\uparrow}^{\alpha'}+J_-c_{k\uparrow}^{\alpha\dagger}c_{k'\downarrow}^{\alpha'}\right.\notag\\
 &  \quad \quad \quad \quad \quad \left.+J_z\left(c_{k\uparrow}^{\alpha\dagger}c_{k'\uparrow}^{\alpha'}-c_{k\downarrow}^{\alpha\dagger}c_{k'\downarrow}^{\alpha'}\right)\right].
 \label{scattering}
\end{align}
In  (\ref{system}) we added the contribution due to an applied or stray magnetic field $\vec{B}$, with $g_J$ being the Land\'{e}-factor, $\mu_B$ the Bohr magneton.
The bath electron, with creation (annihilation) operators  
$c_{k\sigma}^{\alpha\dagger}$ ($c_{k\sigma}^{\alpha}$), have energies $\varepsilon^\alpha_{k \sigma}$ where $\alpha$ can be $T$ or $B$ for the spin-polarized tip or the bulk substrate.
We also account for the voltage of the baths $U^\alpha$.
In  (\ref{scattering})  we concentrate on the interaction of the spins of the electrons with the angular momentum $\vec{J}$ of the atom, which leads to scattering and tunneling processes, with or without spin flip,
with amplitudes $t_{kk'}^{\alpha\alpha'}$ described by the three terms in $H_C$.
For $\alpha=\alpha'=B$ these terms describe the scattering of a bulk electron, while for  $\alpha\neq\alpha'$ they describe tunneling between the tip and the bulk via the Ho atom. Other couplings of the total angular momentum to, e.g., phonons or the radiation field and the resulting dissipative effects are not considered explicitly in this work, but will be accounted for in a qualitative way in section \ref{cNoise}. We further have to keep in mind that the measured current between tip and substrate is also caused by further spin-independent coupling terms.

The eigenstates $|\psi^{\pm}_{m}\rangle$ of the crystal-field Hamiltonian have the property that two eigenstates with the same energy but from different families have vanishing matrix elements $\braket{\psi^{\sigma}_m|J_\nu|\psi^{-\sigma}_{-m}}=0$ for $\nu\in\{+,-,z\}$, $\sigma=\pm$ and all $m$. It is a consequence of two properties of the system: First, the $C_{3v}$ symmetry of the adsorption site, which leads to the Stevens-operators introduced above with operators $J_+$ and $J_-$ appearing in powers of multiples of three. Second, the time-reversal symmetry for $\vec{B}=0$ with the following properties of the time reversal operator $\mathcal{T}$ (see Ref.~\cite{nature12759}),
\begin{align}
  \mathcal{T}^2=&\, 1, \quad \quad \quad \quad \braket{\chi|\phi}=\braket{\mathcal{T}\phi|\mathcal{T}\chi},
\notag\\
  \quad 
\mathcal{T}J_\nu=&-J_\nu\mathcal{T},\quad \mathcal{T}\ket{\psi_m^\sigma}=\ket{\psi_{-m}^{-\sigma}}.
\end{align}
From these relations we find $\braket{\psi_m^\sigma|J_\nu|\psi_{-m}^{-\sigma}}=0$ since 
\begin{align}
  \braket{\psi_m^\sigma|J_\nu|\psi_{-m}^{-\sigma}}=&\braket{\mathcal{T}\psi_{m}^{\sigma}|\mathcal{T}J_\nu\psi_{-m}^{-\sigma}}^{*}\notag=-\braket{\psi_{-m}^{-\sigma}|J_\nu\mathcal{T}\psi_{-m}^{-\sigma}}^{*}\notag\\
  =&-\braket{\psi_{-m}^{-\sigma}|J_\nu|\psi_{m}^{\sigma}}^{*}=-\braket{\psi_{m}^{\sigma}|J_\nu|\psi_{-m}^{-\sigma}}.
\end{align}
Thus, a transition between the two ground states with $m=\pm 8$ 
(and similar $m=\pm 7$) cannot be induced by a single electron scattering, 
which is crucial for their observed long lifetimes \cite{nature12759}. 

On the other hand, at non-zero temperature $T\neq0$ and under the influence of an applied voltage $U$, scattering and tunneling of electrons may lead to transitions to exited states and eventually to transitions between the two ground states. Additionally, time-reversal symmetry breaking terms in the Hamiltonian, such as a magnetic field, give rise to direct transitions between the ground states.  These effects lead to a finite relaxation time $T_1$, which has been probed in the experiment and is the first and central quantity to be studied in this paper. In a later section we will also investigate the decoherence time $T_2$, i.e., the  time scale for the decay of a coherent superposition of the two ground states.

The temperature in the experiments \cite{nature12759} was as low as $0.7 \,\text{K}$ (i.e., $k_BT\approx0.060\,\text{meV}$), whereas the typical system energy, i.e. the first excitation energy is $7.7\,\text{meV}$. Thus, the system is clearly in the quantum regime. 

\section{Quantum master equation} 
\subsection{Damping due to electron scattering and tunneling}
The description of the system in the quantum regime under the influence of the electronic reservoirs requires solving the reduced quantum master equation appropriate for open quantum systems \cite{bStatMethQO,SWB-107920557}, 
\begin{align}
 \dot{\rho}(t)=i\left[\rho(t),H_S\right]+\int_{t_0}^t \text{d}t' \, \Sigma(t-t')\rho(t')\, .
 \label{eME}
\end{align}
Here and below we set $\hbar=1$.
The influence of the two electronic reservoirs enters in the dissipative kernel $\Sigma(t-t')$. We assume it to be of a
Lindblad form and use a Born-Markov approximation. In the interaction picture the quantum master equation for the 17 states of the system then reduces to
\begin{align}
  \label{eMaME}
  \dot{\rho}&_I=\, -\sum_{\substack{\nu,\nu'=+,-,z\\ \alpha,\alpha'=T,B}}\int_{0}^\infty \text{d}t'\\
  &\left\{\vphantom{\int}\right.\left[J_\nu(t)J_{\nu'}(t')\rho_I(t)-J_{\nu'}(t')\rho_I(t)J_{\nu}(t)\right]C^{\alpha\alpha'}_{\nu\nu'}(t-t')\notag\\
  &+\left[\rho_I(t)J_{\nu'}(t')J_{\nu}(t)-J_{\nu}(t)\rho_I(t)J_{\nu'}(t')\right]C^{\alpha\alpha'}_{\nu\nu'}(t'-t)\left.\vphantom{\int}
\right\}, \notag 
\end{align}
with the dissipative kernel expressed by the correlation functions 
\begin{align}
  C^{\alpha\alpha'}_{\nu\nu'}(t)=\sum_{k,k'}|t^{\alpha\alpha'}_{kk'}|^2 \langle s^{\alpha\alpha'}_{kk'\nu}(t) s^{\alpha'\alpha}_{k'k\nu'}(0)\rangle
  \end{align}
  with 
\begin{align}
 s_{kk'-}^{\alpha\alpha'}=&\, c_{k\downarrow}^{\alpha\dagger}c_{k'\uparrow}^{\alpha'},\,\,s_{kk'+}^{\alpha\alpha'}=c_{k\uparrow}^{\alpha\dagger}c_{k'\downarrow}^{\alpha'},\nonumber\\ 
s_{kk'z}^{\alpha\alpha'}=&\, c_{k\uparrow}^{\alpha\dagger}c_{k'\uparrow}^{\alpha'}-c_{k\downarrow}^{\alpha\dagger}c_{k'\downarrow}^{\alpha'}. 
\end{align}
Assuming $t_{kk'}^{\alpha\alpha'}\approx t^{\alpha\alpha'}$
and introducing the spin-dependent electron densities of states at the Fermi-edge $N_{\sigma}^{\alpha}$ with $\sigma= \,\uparrow,\downarrow$ we obtain the Fourier transformed of the correlation functions, 
\begin{align}
 \tilde{C}^{\alpha\alpha'}_{+-}(\Lambda_{nm})=&\,|t^{\alpha\alpha'}|^2N_{\uparrow}^{\alpha}N_{\downarrow}^{\alpha'}\zeta(\Lambda_{nm}+e(U^\alpha-U^{\alpha'})) \nonumber\\
 \tilde{C}^{\alpha\alpha'}_{-+}(\Lambda_{nm})=&\,|t^{\alpha\alpha'}|^2N_{\downarrow}^{\alpha}N_{\uparrow}^{\alpha'}\zeta(\Lambda_{nm}+e(U^\alpha-U^{\alpha'}))\nonumber \\
 \tilde{C}^{\alpha\alpha'}_{zz}(\Lambda_{nm})=&\,|t^{\alpha\alpha'}|^2\left(N_{\uparrow}^{\alpha}N_{\uparrow}^{\alpha'}+N_{\downarrow}^{\alpha}N_{\downarrow}^{\alpha'}\right)\notag\\
 &\times \zeta(\Lambda_{nm}+e(U^\alpha-U^{\alpha'})).
\end{align}
They are evaluated at the energy differences of the atomic system, $\Lambda_{nm}=E_m-E_n$ shifted by the applied voltages. Here we introduced
 \begin{align}
 \zeta(\omega)=\int f(E)\left[1-f(E-\omega)\right]dE=\frac{\omega}{\exp[\omega/(k_BT)] - 1}, \notag
\end{align}
where $f(E)=[e^{E/(k_B T)}+1]^{-1}$ is the Fermi function. %We set the bulk potential $U^B$ to zero. 

As usual in the context of the tunneling magneto-resistance we define the tip polarization $\eta=(P_{\uparrow}-P_{\downarrow})/(P_{\uparrow}+P_{\downarrow})\in[-1,1]$, where the 
spin up/down populations are proportional to the densities of states  $P_{\uparrow/\downarrow}\propto N_{\uparrow/\downarrow}^T=N^T\cdot\tfrac{1}{2}(1\pm \eta)$. The bulk electrode is assumed to be non-polarized,
hence $N^B_{\uparrow/\downarrow}=N^B$.  
The remaining parameters, apart from the polarization $\eta$, can be lumped in the coefficients 
 \begin{align}
c_{\alpha\alpha'}=\frac{1}{2}|t^{\alpha\alpha'}|^2N^\alpha N^{\alpha'}.
\notag
\end{align} 

In the following discussions of the relaxation processes we will concentrate mostly on  applied voltages $U=U^T-U^B$ of the order of or exceeding $3\,\text{mV}$. In this case the effect of tunneling electrons is stronger than that of scattering electrons in the bulk electrode, which in turn is assumed to be stronger than the scattering in the tip. We therefore set $c_{TT}=0$ in most of the paper. Furthermore, we start with $c_{BB}=0$, but we will analyze the influence of bulk electrons scattering in section \ref{cBulk}. 
Also for the calculation of the dephasing time $T_2$ for zero current in section \ref{cT2} the scattering terms need to be taken into account.

\subsection{The proper basis states and environment-induced superselection}
\begin{figure}[htb]
 \includegraphics[origin=c,width=0.48\textwidth]{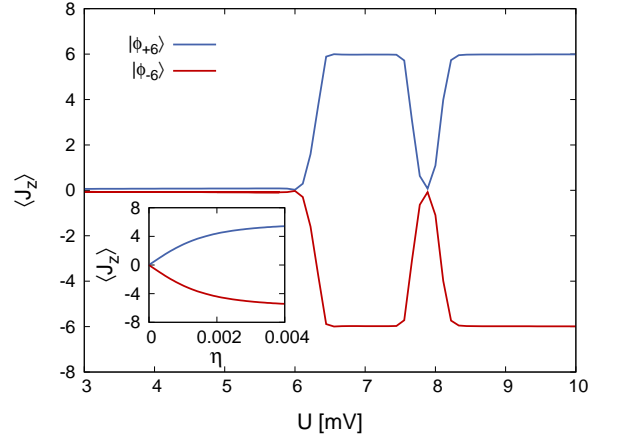}
 \caption{Expectation value $\langle J_z \rangle$ versus the voltage $U$ for a tip polarization $\eta=-0.15$, and in the inset versus the polarization $\eta$ for  $U=7.3\,\text{mV}$. The other parameters are $B_z=10^{-8}\,\text{T}$ and $c_{TB}=3.41\cdot10^{6} (\text{meV s})^{-1}$.}
  \label{pUEtaEinselection}
\end{figure}

If the system behaves sufficiently classically, it is possible to reduce the quantum master equation to rate equations for the populations $P_i$ of the different states
 \begin{align}
   \frac{\mathrm{d} P_{i}}{\mathrm{d} t} = \sum_j \left( \Gamma_{ji} P_{j} - \Gamma_{ij} P_{i} \right),
   \label{eRate}
\end{align}
with $\Gamma_{ji}$ describing the transition rates between the different states. Such a reduction is possible if the coherences, i.e., the off-diagonal components of the density matrix 
decay much faster than non-equilibrium populations. The question remains, what is the appropriate basis to decouple the coherences and populations. One frequently used option for this basis are the eigenstates of $H_{CF}$. 
It is the appropriate choice if the influence of the baths is weak, and the eigenstates are only little affected by their presence. This choice and the resulting reduction to rate equations has been used, e.g., in Refs.\cite{Science-2007-Hirjibehedin-1199-203,nl901066a, PhysRevB.90.155134,PhysRevLett.104.026601,nphys1616} and allows an easy interpretation of transition channels. 

In general, the quantum master equation~(\ref{eMaME}) for the density matrix can be rewritten in the form $\dot{\vec{\rho}}(t)=\mathcal{M}\vec{\rho}(t)$ where  the density {\sl matrix} $\rho$ (here $17 \times 17$) is arranged as a {\sl vector} $\vec{\rho}$ (with 289 components) and all the dynamics, coherent and dissipative, are included in the components of the large 
($289\times 289$) matrix $\mathcal{M}$ (for details see Appendix). The spectral decomposition of this matrix,
\begin{align}
\mathcal{M}\, \vec{\rho}_n=m_n\, \vec{\rho}_n,    
\label{QME}
\end{align}
gives access to various relaxation rates. It also yields
the steady-state populations 
$\vec{\rho}(t\rightarrow \infty)=\vec{\rho}_{st}=\vec{\rho}_0$, which is the eigenvector corresponding to the zero eigenvalue $m_0=0$ \cite{1208.1161}.

If the effect of the baths is not weak a reduction to a classical rate equation is not allowed.
Instead we should study the eigenvalue problem of Eq.~(\ref{QME}).
Its eigenvectors are superpositions of different eigenstates of  $H_{CF}$. This is in particular true for the $0$-family. We recall that due to degeneracies the eigenstates of  $H_{CF}$ corresponding, e.g., to $m=\pm6$ are the symmetric and antisymmetric coherent superpositions of (predominantly) the two $J_z$ basis states, resulting in a vanishing expectation value $\langle J_z\rangle$ displayed in Fig.~\ref{pLevels}~a). Due to the dissipation this coherence may get destroyed. This scenario has been termed ``environment-induced superselection" principle  by Zurek \cite{ProgZurek-281-312,RevModPhys.75.715}.
Decoherence selects favored ``pointer" states which are stable under the influence of the 
environment \cite{1412.5206}. These states are the steady-state eigenstates of the full quantum master equation which we will denote as $\ket{\phi_m}$ with $m=-J,-J+1,\ldots,J$.  If the dissipation is strong these eigenstates are actually much closer to the original $J_z$ basis states with non-vanishing expectation values 
$\langle J_z\rangle\approx \pm 6$. Similarly the states corresponding to $m=\pm3$ get modified. The situation with strong dissipation is depicted in  Fig.~\ref{pLevels}~b). 

It is instructive to study the transition between the two limiting cases.
We focus on the two stationary states of the doublet $\ket{\phi_{+6}}$ and $\ket{\phi_{-6}}$ and calculate their $J_z$ expectation value. The result is depicted in 
Fig.~\ref{pUEtaEinselection}. 
For weak dissipation, e.g., for small $U$ the states reduce approximately to the $H_{CF}$ eigenstates $\ket{\psi^{0}_{6s}}$ and $\ket{\psi^{0}_{6a}}$ with  $\langle J_z\rangle \approx 0$, while for strong dissipation they  approach the states with maximum expectation values $\langle J_z\rangle\approx \pm 6$.
For low voltages the electrons do not have enough energy to scatter into higher excited states and  the superposition states remains stable. For higher voltages exceeding $\sim6\,\text{mV}$ and nonzero temperature excitations are possible and the superposition is destroyed. 
%The environment selects stable states in consideration of scattering of the electrons. When the states $\ket{\psi^{St}_{6+}}$ and $\ket{\psi^{St}_{6-}}$ reach their maximum $J_z$ expectation value of $\sim 6$, they correspond to the states which we former introduced as $\ket{\psi^0_{+6}}$ and $\ket{\psi^0_{-6}}$. 
In Fig.~\ref{pUEtaEinselection} we further note a special feature, namely a dip at $U=7.7\,\text{mV}$, since at this voltage the electrons are in resonance with the first excitation energy. 

In the inset of Fig.~\ref{pUEtaEinselection} the dependence of the environment-induced superselection on the polarization of the tip is shown. Without polarization, the states have no preferred basis in which they evolve upon scattering of the electrons. But already a very small polarization in z-direction (mind the scale of the axes) is enough to drive the system to the pointer states with maximum $\langle J_z \rangle$. 

The environment-induced superselection strongly influences the relaxation processes, which will be studied in the following section.

\section{Relaxation time $T_1$}
If the system is mainly in one of the two degenerate ground states, the relaxation time $T_1$ towards the steady state is given by the smallest non-zero eigenvalue of the matrix $\mathcal{M}$ (see Eq.~(\ref{QME})), corresponding to the eigenvector $\vec{\rho}_1=(1,0,....,0,-1)^T$. (Here we assumed an ordering such that the first and last entries of the eigenvector are the ground states). The inverse of the $T_1$ time, $1/T_1=-m_1=\Gamma_{-8\rightarrow+8}+\Gamma_{+8\rightarrow-8}$, accounts for all relaxation channels from one ground state to the other. Note that the switching rate $\Gamma_{-8\rightarrow+8}$  from $\ket{\phi_{-8}}$ to $\ket{\phi_{8}}$ accounts for the direct transition but also all those via excited states. It thus differs from the rate $\Gamma_{-88}$ of the rate equations (\ref{eRate}), which describes only the direct transition.
The relation between the two switching rates $\Gamma_{-8\rightarrow+8}$ and $\Gamma_{+8\rightarrow-8}$ (and related lifetimes) follows from the steady state populations $P_{-8}$ and $P_{+8}$ of the two ground states, $\tau_{-8}/\tau_{+8}= \Gamma_{+8\rightarrow-8}/\Gamma_{-8\rightarrow+8}=P_{-8}/P_{+8}$. \\

\subsection{Voltage dependence}
To compare with the experiments of Ref.\ \cite{nature12759} we need to determine the coupling strength for tunneling of the reservoir electrons via the Ho atom.
For this purpose we calculate the current $I_{Th}$ from the dissipative part of the master equation (for details see Appendix). By comparing with the experiment we should be able to determine the coupling strength $c_{TB}$. However, the current $I_{Th}$ describes only the current where electrons scatter due to the spin -- angular momentum interaction. The total current,
which in the experiments was always kept at $I_{Exp}=1\,\text{nA}$,
 includes a `leakage' current $I_{Exp}=I_{Th}+I_{Leak}$. It can arise due to electrons tunneling directly between tip and bulk or due to electrons scattering with shells other than the $4f$ shell which forms the basis of the considered angular momentum states. By comparing the current which involves spin flips, and accordingly depends on the the spin state of the Ho atom, with the state-independent current we get a rough estimate. For the following discussion we assume that $I_{Th}$ amounts for roughly 10\% of the total current $I_{Exp}$, i.e. $I_{Th}=0.1\,\text{nA}$. For $U=3\,\text{mV}$ this is obtained for a coupling strength of $c_{TB}=3.41\cdot10^{6} (\text{meV s})^{-1}$.
\begin{figure}[htb]
 \includegraphics[origin=c,width=0.48\textwidth]{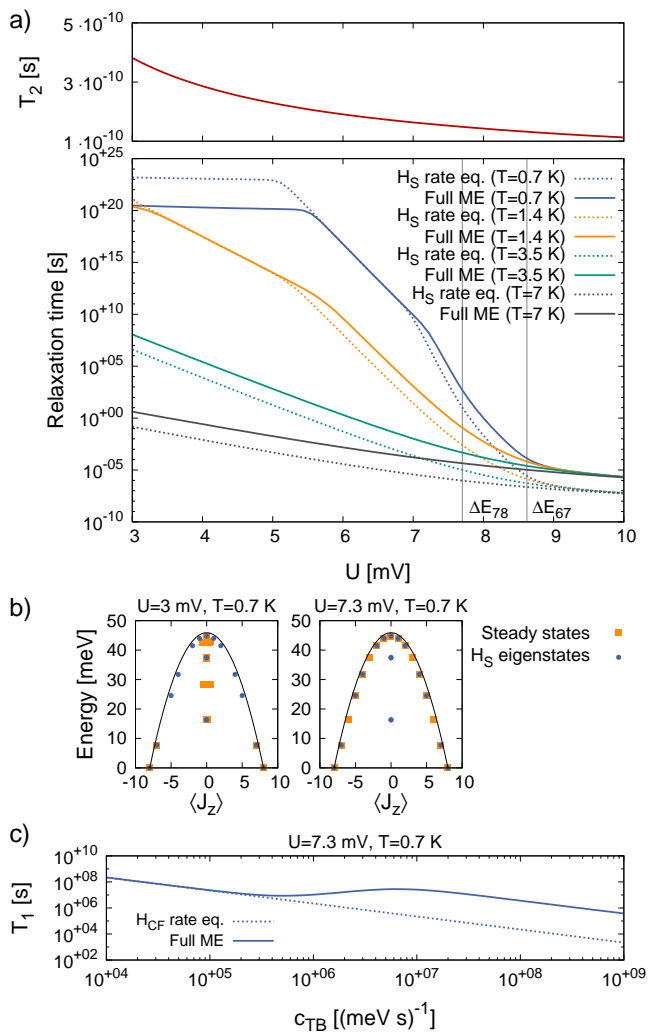}
 \caption{a) Relaxation time $T_1$ and decoherence time $T_2$ of the two ground states versus the applied voltage $U$ obtained from the full quantum master equation (\ref{QME})(full lines) and the rate equation (\ref{eRate}) based on $H_{CF}$  eigenstates (dotted lines). A weak magnetic field $B_z=10^{-8}\,\text{T}$ is applied, the temperatures are $T=0.7\,\text{K}$, $T=1.4\,\text{K}$, $T=3.5\,\text{K}$ and  $T=7\,\text{K}$, and  the coupling strength is $c_{TB}=3.41\cdot10^{6} (\text{meV s})^{-1}$. b) Energy expectation values versus the expectation values $\langle J_z \rangle$  for the steady states of the full quantum master equation and for the $H_{CF}$ eigenstates at  $T=0.7\,\text{K}$ for $U=3\,\text{mV}$ and $U=7.3\,\text{mV}$. c) Relaxation time $T_1$ versus the coupling strength $c_{TB}$ as obtained from the full quantum master equation and the rate equation for $T=0.7\,\text{K}$ and $U=7.3\,\text{mV}$.}
  \label{pLifetUMerge}
\end{figure}

We are now ready to analyze the dependence of the relaxation time on the applied voltage.
In Fig.~\ref{pLifetUMerge} a) we plot the results for the temperature $T=0.7\,\text{K}$ chosen in the experiments, and for comparison also at several higher temperatures $T=1.4 \,\text{K}$, $T=3.5 \,\text{K}$, and $T=7 \,\text{K}$. We compare the $T_1$ time, as obtained from the numerical solution of the full quantum master equation, and the result obtained from the approximate rate equations (\ref{eRate}) in the basis of $H_{CF}$ eigenstates. Similarly we compare in Figs.~\ref{pLifetUMerge}~b) the $J_z$ expectation values of the steady state solutions of the master equation and those of the $H_{CF}$ eigenstates for two 
different values of $U$. For reasons of numerical stability we include in all our simulations the effect of a very weak magnetic field applied in z-direction ($B_z=10^{-8}\,\text{T}$). Otherwise the two ground states get completely decoupled within our numerical precision, and divergences appear, or the reduced density matrix is no longer positive semidefinite. 

Focusing on low temperature, $T=0.7K$, we note that for low voltages, $U \lesssim 3\,\text{mV}$, the six  lowest lying steady states 
of the full quantum master equation have very similar properties as the $H_{CF}$ eigenstates. In this regime transitions between the two ground states $\ket{\psi_{8}^+}$ to $\ket{\psi_{-8}^-}$  are caused mostly by the (weak) symmetry-breaking  magnetic field and are thus voltage independent.  
The full master equation yields shorter lifetimes than the rate equation in this regime, because of additional coherent transitions contained in the full theory.

At higher voltages we observe for $T=0.7\,\text{K}$ and $T=1.4\,\text{K}$ in (the semi-log plot of) Fig.\ref{pLifetUMerge} a) an exponentially activated behavior. In this regime the full master equation and the rate equation yield very similar results. Here
the main switching channel is via the first excited states, $\ket{\phi_{7}}$ or $\ket{\phi_{-7}}$, followed by a fast decay to the other ground state on the other side of the parabola. 
Since the first step is the bottleneck of the process we have $1/T_1\approx\Gamma_{87}+\Gamma_{-8-7}$. 
For an estimate we ignore the effect of the tip polarization and of a magnetic field (i.e. $\Gamma_{87}
\approx \Gamma_{-8-7}$) and get 
\begin{align}
\Gamma_{87}=& \, c_{TB}|\braket{\psi_7^{-}|J_-|\psi_8^+}|^2\cdot\zeta(E_7-E_8-eU)\\
\approx& \, 16c_{TB}\cdot\frac{(E_7-E_8-eU)}{e^{(E_7-E_8-eU)/k_BT}-1}
\end{align}

For still higher voltages,  $U>7\,\text{mV}$,  the results obtained in the two approaches differ again. In this regime, in the frame of the rate equation excitations to the symmetric and anti-symmetric $H_{CF}$-eigenstates $\ket{\psi^0_{6s}}$ and $\ket{\psi^0_{6a}}$ become possible. These two states have high transition rates between each other and thus provide a `shortcut' for the decay. However, in the full master equation these superpositions states are replaced
by the steady states $\ket{\phi_{+6}}$ and $\ket{\phi_{-6}}$, with properties 
illustrated in Figs. \ref{pUEtaEinselection} and \ref{pLifetUMerge} b), which in this parameter regime are actually close to $\langle J_z \rangle=\pm 6$ states. They are weakly coupled, and the shortcut is no longer open, which increases the relaxation time. 

For even higher voltages,  $U > 8.6\,\text{mV}$, tunneling electrons have enough energy to overcome also the second energy excitation gap of $\Delta E_{67}=E_6-E_7\approx8.6\,\text{meV}$, which is the largest gap of the system. From this point on, sequential scattering over the top of the parabola is the main transition channel, and the slope of $T_1(U)$ versus $U$ changes.

At higher temperatures, the different regimes get smeared out, as can be seen in the plot of 
Fig.~\ref{pLifetUMerge}, especially for $T=3.5\,\text{K}$ and $T=7\,\text{K}$. In this regime the main transition channel is always via higher excited states. For voltages larger than $U\approx9\,\text{mV}$ the results for all temperatures are very similar, because in all cases most of the electron scattering leads to the transition over the top of the parabola. Again, the rate equations overestimate the role of transitions via the 'shortcut' states $\ket{\psi^0_{6s}}$, $\ket{\psi^0_{6a}}$, $\ket{\psi^0_{3s}}$ and $\ket{\psi^0_{3a}}$.\\

In Fig.~\ref{pLifetUMerge}~c) we investigate the dependence of the lifetime on the coupling strength $c_{TB}$ for low temperature and $U=7.3\,\text{mV}$. From the rate equations we find simply that  $T_1$ time decreases proportional to $1/c_{TB}$. But the solution of the full quantum master equation yields different results.  
In the considered regime
excitations to the $\ket{\phi_{+6}}$ and $\ket{\phi_{-6}}$ play a role. The coherence leading to these superposition states is increasingly destroyed with growing coupling strength. As a result in the range  $10^5(\text{meV s})^{-1}\le c_{TB} \le 10^7(\text{meV s})^{-1}$  the stronger coupling even stabilizes the system by decoupling the two sides of the parabola.

\subsection{Further details of the experiments}
\label{cExpU}
\begin{figure*}[htb]
 \includegraphics[origin=c,width=0.98\textwidth]{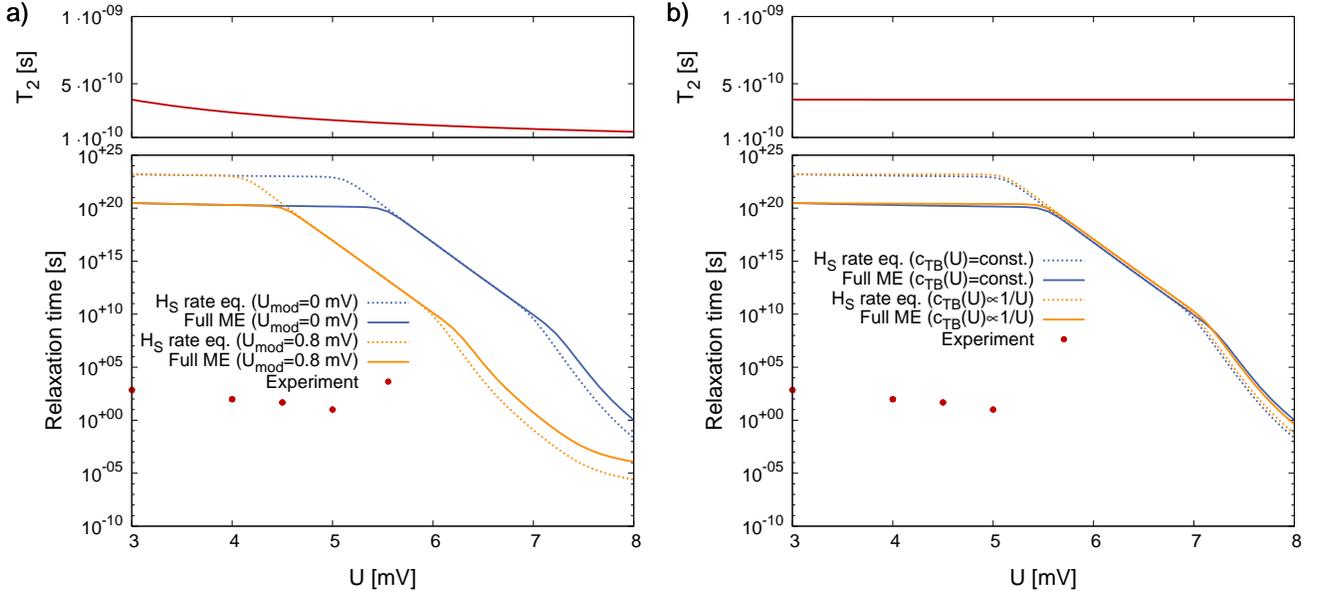}
 \caption{Relaxation time $T_1$ and decoherence time $T_2$  of the two ground states versus the applied voltage $U$  in a weak magnetic field $B_z=10^{-8}\,\text{T}$ at $T=0.7\,\text{K}$ for the full master equation solution and the rate equation (\ref{eRate}) with the $H_{CF}$ eigenstates. a) With and without modulation voltage $U_{mod}=0.8\,\text{mV}$. b) With and without tip distance correction. For comparison the experimental data are shown in red. The error bars are too small to be visible.}
  \label{pLifetUMerge_2}
\end{figure*}
To compare with the experiments of Ref.\ \cite{nature12759}, two further effects have to be taken into account. First, in the experiment the applied voltage was modulated, $U_{tot}(t)=U+\sqrt{2}U_{mod}\cos(\omega t)$, with  $U_{mod}=0.8\,\text{mV}$.
The frequency of the modulation,  $\omega=720\,\text{Hz}$, is several orders of magnitude faster than the timescales of interest. We model this fast oscillation by suitably averaging the correlation functions with the distribution $h(U')=1/\pi\left[2 (eU_{mod})^2-(eU')^2\right]^{-1/2}$ for $|U'|\leq\sqrt{2}U_{mod}$ and $h(U')=0$ otherwise \cite{1973_Klein_et_al}. This means, e.g.
\begin{align}
  \tilde{C}^{TB}_{+-}(\Lambda_{nm})&= \,c_{TB}\frac{1}{2}(1+\eta)\notag\\
   \times &\int \text{d}U'\zeta(\Lambda_{nm}+eU^T+eU')h(U').
\end{align}
The effect on the lifetimes is shown in Fig.\ \ref{pLifetUMerge_2} a). In essence, the modulation amounts to a shift $U\rightarrow U + \sqrt{2}\cdot U_{mod}$, by approximately $1.1\text{mV}$.

Second, in the experiment the voltage-induced tunnel current was measured, and in fact, by adjusting the distances between tip and electrode, was kept fixed at the value $I_{Exp}=1\,\text{nA}$. 
To keep the current fixed for different voltages $U$, we allow
in the simulations $c_{TB}(U)$ to be voltage dependent. We checked numerically that the tunnel 
contact behaves close to Ohmic,  $I_{Th}\approx U/R_{Tun}$. We therefore adjust the coupling 
$c_{TB}(U)\propto 1/U$ to keep the current constant. We recall that an estimated 90\% of the current is leakage current. With $c_{TB}(U=3\,\text{mV})=3.41\cdot10^{6} (\text{meV s})^{-1}$ we arrive at 
$I_{Th}=0.1\,\text{nA}$. We will proceed using these values in all simulations reported below. 
All this being said and done, we note that the effect of the adjustment, which is included in Fig.~\ref{pLifetUMerge_2}~b), is weak.

\section{Deviations from the ideal situation} 

When comparing the calculated $T_1$ times with the experimental data, as illustrated in Figs.\ \ref{pLifetUMerge_2}, we note that the theory produces far too long times.
Therefore, we need to take a closer look at the experiment and possible deviations from the ideal situation assumed so far. 

\subsection{Scattering of bulk electrons}
\label{cBulk}
\begin{figure*}[htb]
 \includegraphics[origin=c,width=0.94\textwidth]{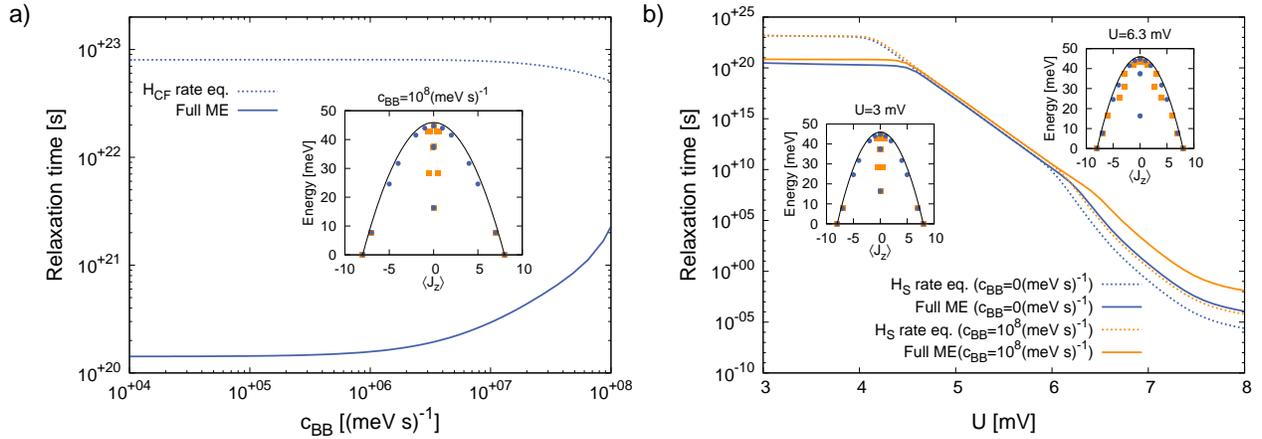}
 \caption{a) Relaxation time $T_1$ versus the bulk electron scattering strength $c_{BB}$ as obtained from the full master equation (\ref{QME}) and the rate equation (\ref{eRate}) based on  $H_{CF}$ eigenstates. The parameters are $U=5\,\text{mV}$, $B_z=10^{-8}\,\text{T}, c_{TB} = 3.41\cdot10^{6} (\text{meV s})^{-1}$. b) Relaxation time $T_1$ versus the voltage $U$ for the model with modulation voltage broadening and tip distance correction with and without bulk electron scattering strength $c_{BB}=1.0\cdot10^7\,(\text{meV s})^{-1}$.}
  \label{pLifetcBB}
\end{figure*}
Up to now we ignored the effect of bulk electrons scattering from the Ho atom,
and the question arises whether it could be the source of the mismatch between theory and experiment.
The scattering processes are easily included in the quantum master equation, and their effects
are illustrated in Fig.~\ref{pLifetcBB}. 
\begin{figure*} [hbt]
 \includegraphics[origin=c,width=0.99\textwidth]{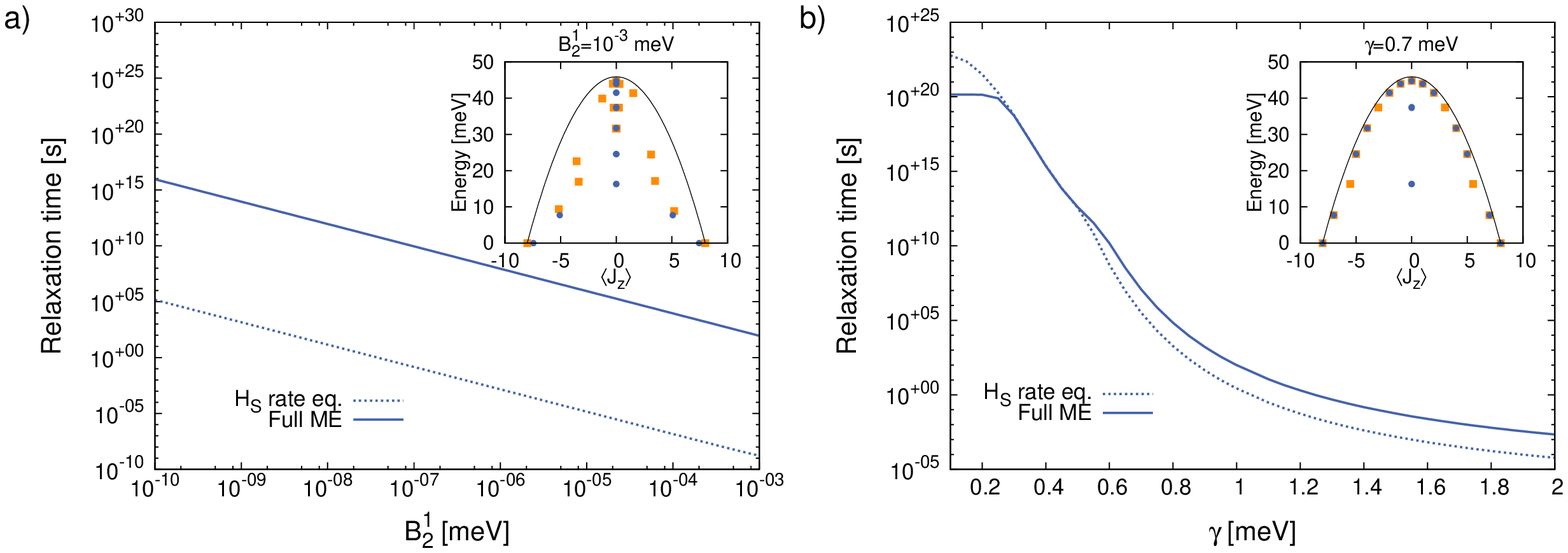}
 \caption{Relaxation time $T_1$ of the two ground states under the influence of deviations from the perfect situation. We compare the results obtained from the full master equation (\ref{QME}) and  the rate equation (\ref{eRate}) based on the $H_{CF}$ eigenstates.  a) $T_1$ versus Stevens parameter $B_2^1$ characterizing the deviation from the perfect trigonal symmetry. b) $T_1$ versus the strength of Gaussian broadening $\gamma$ chosen to account for further lifetime broadening effects. In both plots we choose the parameters $U=5\,\text{mV}$ and $B_z=10^{-8}\,\text{T}$. The insets show the $J_z$ expectation value of the steady states and the $H_{CF}$ eigenstates.}
  \label{pLifetBxMerge}
\end{figure*}
In the first panel we show how the relaxation time depends on the coupling strength $c_{BB}$ for a fixed value of the tunneling strength $c_{TB}=3.41\cdot10^{6} (\text{meV s})^{-1}$ and voltage $U=5\,\text{mV}$. As long as $c_{BB}$ is smaller than $c_{TB}$ the lifetime remains nearly unchanged. For stronger  $c_{BB}$, the scattering of bulk electrons leads to suppression of coherent transitions and thus to {\sl longer} lifetimes. This arises because of the low temperature of the bulk electrodes, which cools the system into the ground states, whereas the tunneling electrons due to the applied voltage have enough energy to excite the system. 
The combination of voltage-dependent tunneling and voltage independent scattering is illustrated in Fig.~\ref{pLifetcBB}~b) (including the effects of modulation voltage broadening and tip distance correction  described above). We assume a bulk electron scattering strength $c_{BB}=1.0\cdot10^7\,(\text{meV s})^{-1}$ which is slightly higher than the tunneling coupling strength $c_{TB}=3.41\cdot10^{6} (\text{meV s})^{-1}$. Again we note the increase of the lifetime as a consequence of the scattering.

\subsection{Breaking the $C_{3v}$-symmetry} 
As an example of a symmetry breaking term, we consider the effect of the $C_{3v}$-symmetry-breaking term of the Stevens operators, $B_2^1\cdot O_2^1=B_2^1\cdot \left[J_z,J_++J_-\right]_+$ \cite{Altshuler1974}.
This term arises if the tip is not perfectly centered over the Ho atom, or if nearby surfaces or imperfections in the crystal break the symmetry. It breaks all the rotational symmetries of the system. A magnetic field in the xy-plane would have a similar effect.

The symmetry-breaking parameter $B_2^1$ is varied in Fig.~\ref{pLifetBxMerge}~a) between $10^{-10}\,\text{meV}$ and $10^{-3}\,\text{meV}$, which is still orders of magnitude lower than the leading crystal-field  term $B_2^0=-0.239\,\text{meV}$.  As a result of the symmetry breaking the eigenstates of $H_{CF}$ get mixed, and the protection against direct transitions is lost. In Fig. \ref{pLifetBxMerge} a), the voltage is chosen to be $U=5\,\text{meV}$, so the leading transition is directly between the two ground states. With rising strength of $B_2^1$ the relaxation time decreases  $T_1\propto (B_2^1)^{-2}$. 
In the frame of the rate equations the switching arises since electron tunneling directly couples the two states. In the frame of the full master equation scattering of electrons destroys the phase coherence of these superpositions and the resulting expectation values $\langle J_z \rangle$ of the steady states are closer to the parabola, characteristic for the $H_{CF}$ eigenbasis (see the ground states of the inset of 
Fig.~\ref{pLifetBxMerge}~a)). This means, that the symmetry protection gets restored and the lifetimes increase drastically. All this depends on the coupling strength $c_{TB}$, which is the important parameter for the superselection (see Fig. \ref{pLifetUMerge} c)). Unfortunately, the parameter $B_2^1$ is not independently accessible in the experiment, and  it is difficult to draw more precise conclusions.

\subsection{Noise in the circuit}
\label{cNoise}
To account for the influence of further perturbations, such as thermal noise in the electronics, we introduce a lifetime broadening. Specifically we average the results obtained so far, assuming a Gaussian broadening of the tunneling electron energy distribution $g(\delta E, \gamma)=1/\sqrt{2\pi \gamma^2}\, \exp[-\delta E^2/(2\gamma^2)]$ with a width characterized by the parameter $\gamma$ \cite{1973_Klein_et_al}. Its influence on the lifetime $T_1$ is visualized in Fig.\ \ref{pLifetBxMerge} b). The relaxation time is strongly reduced by this broadening  because the broadening allows excitations into one of the $\ket{\phi_{7}}$ or $\ket{\phi_{-7}}$ states, followed by a subsequent decay to the other ground state. 

\begin{figure}[htb]
 \includegraphics[origin=c,width=0.48\textwidth]{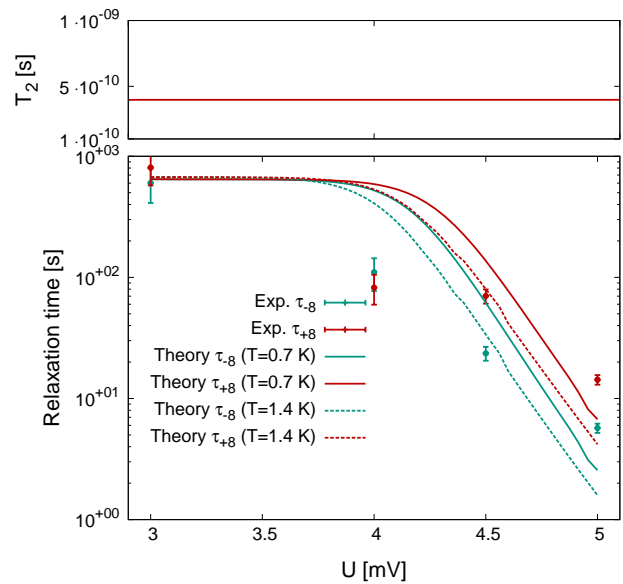}
 \caption{Relaxation time $T_1$ and decoherence time $T_2$ of the two ground states versus the applied voltage $U$ at $T=0.7\,\text{K}$ and $T=1.4\,\text{K}$ for the  modulation amplitude $U_{mod}=0.8\,\text{mV}$, lifetime broadening $\gamma=0.95\,\text{meV}$, magnetic field $B_z=1\cdot10^{-8}\,\text{T}$, symmetry breaking $B_2^1=4\cdot10^{-4}\,\text{meV}$, and tip spin polarization $\eta=0.15$. 
 We compare the results obtained from the full master equation (\ref{QME}) (solid lines) and  the rate equation (\ref{eRate}) based on the $H_{CF}$ eigenstates (dashed lines).
The theoretical values are compared with the state-dependent data from the experiment with error bars indicating the statistical errors of the measurement.}
  \label{pLifetUExp}
\end{figure}

By fitting the parameters characterizing the two deviations from the ideal situation we manage 
to obtain results for the
$T_1$ times close to the the experimental ones (see Fig.~\ref{pLifetUExp}). We plot results for two values of the temperature $T=0.7\,\text{K}$ and $T=1.4\,\text{K}$, where the first one is the bath temperature in the experiment, while the second is chosen to account for some electron heating induced by the current. (An increased temperature alone would not be sufficient to explain the discrepancy between the simulations and the experiment.) Although with the fitting we reached a reasonable agreement with experiment we have to acknowledge that the fit is not conclusive, since the number of data points is too low to determine the parameters independently.

\subsection{Magnetic field dependence} 
\begin{figure*}[hbt]
 \includegraphics[origin=c,width=0.99\textwidth]{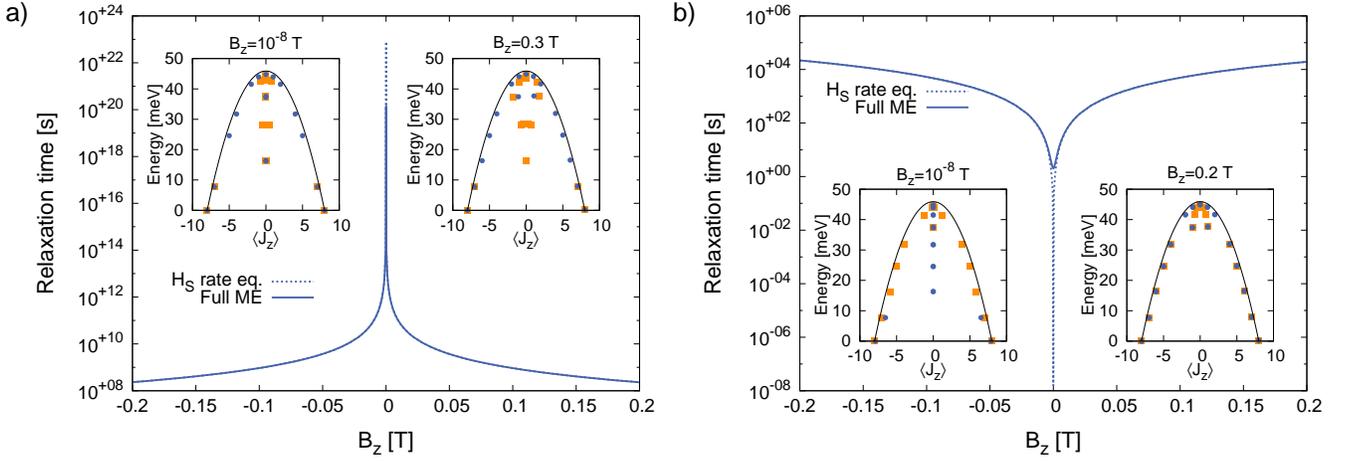}
 \caption{Relaxation time $T_1$ of the two ground states versus a magnetic field $B_z$ as obtained from the full master equation (\ref{QME}) and the rate equations (\ref{eRate}) based on $H_{CF}$ eigenstates. The insets show the  expectation value of $J_z$ for the two descriptions. a) Results for the ideal model for $U=5\,\text{mV}$. b) Results with symmetry-breaking and Gaussian broadening corresponding to the fits of Fig. \ref{pLifetUExp}, i.e., $U_{mod}=0.8\,\text{mV}$, $\gamma=0.95\,\text{meV}$, and $B_2^1=4\cdot10^{-4}\,\text{meV}$.}
  \label{pLifetBzMerge}
\end{figure*}
Next, we investigate the effect of an applied or stray magnetic field $B_z$ which is probably present in the experiments. Its influence depends strongly on the values of the other parameters. In Fig.\ \ref{pLifetBzMerge} a) we show the resulting modification 
for the regime where the main transition, although with small rate, is directly between the states 
$\ket{\psi_{8}^+}$ and $\ket{\psi_{-8}^-}$. We choose $U=5\,\text{mV}$, i.e., we are still in the  regime of voltage-independent relaxation time of Fig.\ \ref{pLifetUMerge} a). As shown in Fig.\ \ref{pLifetBzMerge} a) the lifetime is strongly reduced by the symmetry breaking magnetic field $B_z$, since the symmetry protection of the two ground states is lost. This behavior is obtained both from the rate equation and the solution of the quantum master equation. 

Interestingly, as illustrated by Fig.\ \ref{pLifetBzMerge} b), the behavior can be completely different for different parameters. In this example the lifetime increases when a field is applied. For the chosen parameters  the $B_z$ field stabilizes the $J_z$ eigenstates and reduces the switching, which otherwise would be induced by the symmetry breaking term $B_2^1$. We conclude that a detailed study of the magnetic field dependence should provide a better understanding of the different perturbations acting on the Ho adatom.

\subsection{Alternative choice of the crystal-field parameters}
\label{cDonati}
\begin{figure}[hbt]
 \includegraphics[origin=c,width=0.48\textwidth]{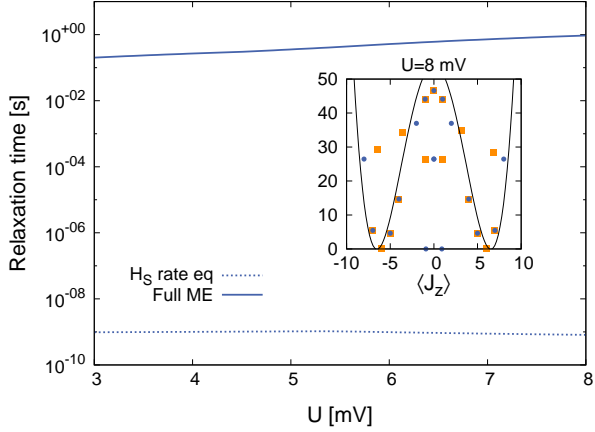}
 \caption{Relaxation time $T_1$ of the two ground states versus the voltage $U$ as obtained from the full master equation and from the rate equations (\ref{eRate}) based on $H_{CF}$ eigenstates for $B_z=10^{-8}\,\text{T}$. The inset shows the  expectation value of $J_z$ for the two descriptions for $U=8\,\text{mV}$. The black line shows the function $f(J_z)=-140\,\mu\text{eV}\cdot O_2^0+1\,\mu\text{eV}\cdot O_4^0+\text{const.}$.}
  \label{pLifetDonati}
\end{figure}
In a recent publication  Donati {\sl et al.} \cite{PhysRevLett.113.237201} reported about x-ray absorption spectroscopy (XAS) and magnetic circular dichroism (XMCD) measurements performed with Ho on Pt(111). The results could be fitted with a crystal-field Hamiltonian of the form (\ref{eHCF}) but with  only the coefficients  $B_2^0=-140\,\mu\text{eV}$ and $B_4^0=1\,\mu\text{eV}$ differing from zero, which differs significantly from what Miyamachi {\sl et al.}~\cite{nature12759} used, and what we assumed so far in this paper. To simulate the lifetimes in our theory with these parameters, we add a small $B_4^3\approx0.3\,\mu\text{eV}$ which we take from Miyamachi {\sl et al.} \cite{nature12759}, because otherwise the ground states decouple completely. With these inputs, we arrive at the level scheme shown in the inset of Fig.~\ref{pLifetDonati}. 
Most important we note that the ground states are now the $\ket{\psi^0_{6s}}$ and $\ket{\psi^0_{6a}}$, which are much stronger coupled (not symmetry-protected) than the states $\ket{\psi_{8}^+}$ and $\ket{\psi_{-8}^-}$ and hence should have much shorter lifetimes.
We analyzed the relaxation rate in the same way as for the other model, with results shown in 
Fig.~\ref{pLifetDonati}. We note that the solution of the rate equation yields very short relaxation times of the order of nanoseconds. On the other hand, the simulation based on the full master equation yields longer lifetimes. The reason is again the environment-induced superselection which destroys the superpositions.
The differences between the two sets of eigenstates are pronounced, as can be seen in the inset of Fig.~\ref{pLifetDonati}. In fact the difference between the two theoretical approaches is even more pronounced than found in the model based on the parameters of  Miyamachi {\sl et al.}.  
We further note from Fig.~\ref{pLifetDonati} that higher voltages even stabilize the ground states, because the excited states couple less to the states on the opposite side of the parabola than $\ket{\psi^0_{+6}}$ and $\ket{\psi^0_{-6}}$ do, and hence the excitation of those states reduces the transition rate. This 
voltage dependence is in stark contrast to the observations made by Miyamachi {\sl et al.} \cite{nature12759}, where the lifetimes decrease with increasing voltages. 

The comparison of theory and experiment on the voltage dependence of the relaxation time supports the Stevens parameters used by Miyamachi {\sl et al.} \cite{nature12759}. 
They were derived by ab initio DFT calculations for a situation where individual Ho atoms were adsorbed on high-symmetry fcc sites on the surface of Pt(111).
In contrast, the experiments of Donati {\sl et al.} \cite{PhysRevLett.113.237201} were performed 
with a high coverage of Ho atoms of $0.04$ monolayers occupying a mixture of hcp and fcc sites.
Further investigations are needed to clarify whether this is the origin of the differing results.

\section{Decoherence time $T_2$} \label{cT2}
The decoherence time $T_2$ is the time scale on which the phase information in a coherent superposition, here specifically of the states $\ket{\psi_{8}^+}$ and $\ket{\psi_{-8}^-}$, gets lost. In the quantum master equation treatment of the problem, $T_2$ is obtained from the decay rate of the off-diagonal matrix element $\rho_{+8-8}$, which is given by the corresponding matrix element in the matrix $\mathcal{M}$,
\begin{align}
1/T_2=& \,- \mathcal{M}_{8-8\rightarrow8-8} \nonumber\\
 \approx& \, 4c_{TB}\braket{\psi_8^+|J_z|\psi_8^+}\braket{\psi_{-8}^-|J_z|\psi_{-8}^-}\zeta(-eU)\nonumber\\
 \approx& \, 264\, c_{TB}\, eU.
\end{align}
The resulting $T_2$ times are plotted for different parameters in Figs.~\ref{pLifetUMerge},~\ref{pLifetUMerge_2} and \ref{pLifetUExp}. It turns out that the $T_2$ time depends mostly on the current, i.e., on the number of scattered electrons. Roughly one can argue that each scattered electron dephases the superposition state, independent of the energy of the electron. Thus, we detect only  in Fig. \ref{pLifetUMerge} a) and Fig. \ref{pLifetUMerge_2} a)  a dependence of $T_2$ on the voltage $U$, because in all other plots the current is kept constant (achieved by the adjustment of $c_{TB}(U)$). We note that the decoherence times $T_2$  are always very short ($\sim10^{-10}\,\text{s}$) making the considered memory unsuitable as a quantum mechanical bit (qubit). Additionally, we found that $T_2$ does not depend on the symmetry breaking $B_2^1$, the broadening $\gamma$, or the magnetic field $B_z$.
If the voltage is set to zero, all terms $\tilde{C}_{\nu\nu'}^{\alpha\alpha'}(\pm\Lambda_{nm})$ are similarly important. Hence $T_2$ is given by
\begin{align}
 1/T_2=& \,- M_{8-8\rightarrow8-8} \nonumber\\
 \approx & \, 4(\sum_{\alpha\alpha'}c_{\alpha\alpha'})\braket{\psi_8^+|J_z|\psi_8^+}\braket{\psi_{-8}^-|J_z|\psi_{-8}^-}\zeta(0)\nonumber\\
 \approx & \, 264 \, k_BT \sum_{\alpha\alpha'}c_{\alpha\alpha'}.
\end{align}
If one assumes that $ \sum_{\alpha\alpha'} c_{\alpha\alpha'}\approx10^6(\text{meV s})^{-1}$ and the temperature is $T=1\,\text{K}$, the decoherence time is roughly $10^{-8}\,\text{s}$.

\section{Initialization} 
We have seen that the relaxation time depends strongly on various parameters. In this section we will demonstrate that by switching parameters suitably we can initialize a specific angular momentum state \cite{PhysRevLett.108.196602}. This enables the writing process if  the Ho atom is used as a memory.
As a specific example we study what happens to the state $\ket{\psi_{-8}^-}$ when we pulse the system for a time $t_p$ with a voltage $U$ and then let the system relax for $1\,\mu\text{s}$ without a voltage applied. After the relaxation, the population of the state $\ket{\psi_{8}^+}$ is measured, which provides the information about the switching probability $S_{-8\rightarrow 8}(U, t_p)$. 

\begin{figure}[hbt]
 \includegraphics[origin=c,width=0.48\textwidth]{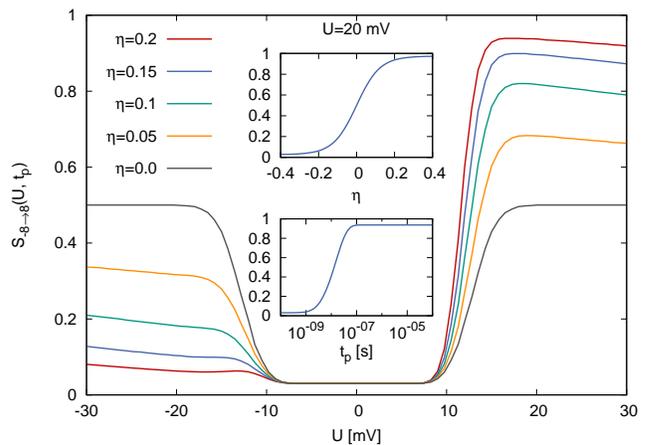}
 \caption{The  probability for switching from $\ket{\psi_{-8}^-}$ to $\ket{\psi_{8}^+}$ versus the applied pulse voltage for different values of the tip polarization $\eta$ and a pulse time of $t_p=2.5\cdot10^{-7}\,\text{s}$. The upper inset shows the switching probability versus the tip polarization  for pulse strength $U=20\,\text{mV}$ and pulse time  $t_p=2.5\cdot10^{-7}\,\text{s}$. The lower inset shows the dependence on the pulse time $t_p$   for pulse strength of $U=20\,\text{mV}$ and $\eta=0.2$.}
  \label{pSwitch}
\end{figure}

In Fig.~\ref{pSwitch}, the switching probability $S_{-8\rightarrow 8}(U, t_p)$ as obtained from the quantum master equation is plotted versus the strength of the applied voltage pulse  for different spin polarizations $\eta$ of the tip. (We neglect again the influence of the bulk electron scattering.)
For $U>20\,\text{mV}$ the value of $S_{-8\rightarrow 8}(U, t_p)$ may get close to 1, i.e., the pulse flips the Ho atom with a high probability into the state $\ket{\psi_{8}^+}$. On the other hand, a pulse with opposite voltage $U<-20\,\text{mV}$  leaves the state with high probability in the state $\ket{\psi_{-8}^-}$. Thus,  both ground states can be prepared by applying voltage pulses with either sign. For voltages between $-10\,\text{mV}\lesssim U\lesssim 10\,\text{mV}$  basically no switching of the Ho spin state is induced,
$S_{-8\rightarrow 8}(U, t_{\rm p})\approx 0$. Around $U=20\,\text{mV}$ the  switching is strongest as long as  $\eta\neq0$. For an unpolarized tip, $\eta=0$, a pulse with high voltage produces a balanced population of the two ground states. 

The upper inset of Fig.~\ref{pSwitch} shows the dependence of the switching probability $S_{-8\rightarrow 8}(U, t_p)$ on the spin polarization $\eta$  for optimum conditions  $U=20\,\text{mV}$ and  pulse time $t_p=2.5\cdot10^{-7}\,\text{s}$. As one could expect, the higher the polarization, the better the initialization. But, the switching between the ground states also requires a certain number of electrons, which increases with longer pulse time $t_p$, as displayed in the lower inset of Fig. \ref{pSwitch}. For the optimal voltage of $U=20\,\text{mV}$ we calculate a current of $I_{Th}\approx0.75\,\text{nA}$. A pulse time of $t_p\approx250\,\text{ns}$ implies then that around 1100 electrons are needed to prepare the atom in one state. By changing the tip distance in the experiment, this dependence could be probed.

Our simulations suggest that it is possible to prepare the system in a required state using pulse times of a few hundreds nanoseconds with high fidelity as long as the tip polarization is large enough. The required pulse time depends strongly on the parameter $c_{TB}$. 
Thus, an experiment with different pulse lengths could help to determine this value.

\section{Conclusion}
\label{cConclusion}
When investigating the dynamics of the angular momentum states of Ho on Pt(111) we found that in an important regime of parameters the system behaves deeply quantum mechanically and cannot be described by a rate equation for transitions between the eigenstates of the crystal-field Hamiltonian $H_{CF}$. Rather the system has to be treated by the full quantum master equation. Its steady-state basis in general differs from the $H_{CF}$ eigenstates, which is an example of the ``environment-induced superselection" 
principle \cite{ProgZurek-281-312,RevModPhys.75.715}.
We analyzed how the relaxation time $T_1$ depends on various parameters of the system. In ideal situations at low temperatures it should be extremely long. 
In contrast the decoherence time $T_2$ is always very short.
We further described a method to initialize the system in one of the two ground states by suitable voltage pulses.

By a proper choice of the parameters describing deviations from the ideal situation we could roughly fit the experiments, with still remarkably long lifetimes. 
Our analysis shows that if the parameters could be improved and controlled more precisely  the system would acquire even longer lifetimes than observed already. 
The system therefore promises to be useful as a single-atom memory  with the possibility to write into the memory by very short pulses of electric currents with high fidelity. 

At this stage, we have to conclude that there are too few experimental data available to determine the parameters independently. 
A further detailed investigation of the coupling parameter $c_{TB}$ and the other parameters is required to identify the main perturbation which limits the lifetime of the Ho adatom in the experiment. 
The lifetime depends strongly on an applied magnetic field.
The analysis of this effect would help obtaining the missing information on the parameters.\\

\section{Acknowledgments}
We thank C. H\"ubner, D. Pfannkuche, S. Andr{\'e}, A. Heimes, S. Zanker, D. Mendler and P. Kotetes for stimulating discussions
and  C. Robach for the illustration in Fig.\ \ref{pSetting}. 

\appendix
\section{Stevens operators and parameters}
For convenience we list here the Stevens operators which are needed to describe the system Ho on Pt(111) 
 \cite{nature12759,Wybourne1965}
\begin{align}
  O_2^0 =& 3J_z^2-J(J+1),\\
  O_4^0 =& 35J_z^4-30J(J+1)J_z^2+25J_z^2-6J(J+1)\notag\\
  &+3J^2(J+1)^2, \\
  O_4^3 =& \frac{1}{4} \left[J_z(J_+^3+J_{-}^3)+(J_+^3+J_{-}^3)J_z \right], \\
  O_6^0 =& 231J_z^6-315J(J+1)J_z^4\notag\\
  &+735J_z^4+105J^2(J+1)^2J_z^2-525J(J+1)J_z^2 \notag\\
  &+294J_z^2-5J^3(J+1)^3+\notag\\
  &40J^2(J+1)^2-60J(J+1), \\
  O_6^3 =& \frac{1}{4} \left[(11J_z^3-3J(J+1)J_z-59J_z)(J_+^3+J_-^3)\right. \notag\\ 
  &+ \left. (J_+^3+J_-^3)(11J_z^3-3J(J+1)J_z-59J_z)\right], \\
  O_6^6 =& \frac{1}{2} \left[ J_+^6+J_-^6 \right].
\end{align}
We also list the parameters which were obtained from ab-initio simulations and listed in Ref. \onlinecite{nature12759}.
\begin{table}[h]
\begin{tabular}{|c|c|}
\hline
Anisotropy constant & value \\ \hline
$B_2^0$ & -239 $\mu$eV \\ \hline
$B_4^0$ &  86 neV \\ \hline
$B_4^3$ & 293 neV \\ \hline
$B_6^0$ & 0.186 neV \\ \hline
$B_6^3$ & -1.967 neV \\ \hline
$B_6^6$ &  0.630 neV \\ \hline
\end{tabular}
\caption{Anisotropy parameters as used in Ref. \onlinecite{nature12759}.}
\end{table}

\section{Setting up the matrix $\mathcal{M}$}
\label{cSupFull}
In this section we present the approach we use to solve the quantum master equation in Born-Markov approximation.
By using the explicit time dependence of the system operators $J_\nu(t)$  in the interaction picture we evaluate the time integrals including the correlation functions of the master equation \cite{PhysRevB.85.174515, 1402.6205}
\begin{align}
 S_{\nu\nu'}^{\alpha\alpha'}(\pm\tau)\equiv \int_0^\infty\text{d}\tau \, C^{\alpha\alpha'}_{\nu\nu'}(\pm \tau)e^{-iH_S\tau}J_je^{iH_S\tau} \, .
\end{align}
In the eigenbasis of $H_S\ket{n}=E_n\ket{n}$, the matrix elements of $S_{\nu\nu'}^{\alpha\alpha'}(\pm\tau)$ become
\begin{align}
 \bra{n}S_{\nu\nu'}^{\alpha\alpha'}\pm&tau)\ket{m}=\int_0^\infty\text{d}\tau \, C^{\alpha\alpha'}_{\nu\nu'}(\pm \tau)\notag\\
  &\bra{n}e^{-iH_S\tau}\ket{n}\bra{n}J_{\nu'}\ket{m}\bra{m}e^{iH_S\tau}\ket{m}\\
  =\bra{n}J_{\nu'}&\ket{m}\int_0^\infty\text{d}\tau \, C^{\alpha\alpha'}_{\nu\nu'}(\pm \tau)e^{i\Lambda_{nm}\tau}\\
  =\bra{n}J_{\nu'}&\ket{m}\left[\frac{1}{2}\tilde{C}^{\alpha\alpha'}_{\nu\nu'}(\pm\Lambda_{nm})-iP\int\frac{\text{d}\omega}{2\pi} \frac{\tilde{C}^{\alpha\alpha'}_{\nu\nu'}(\omega)}{\Lambda_{nm}\mp\omega}\right].
\end{align}
Here, $\Lambda_{nm}=E_m-E_n$ are the energy differences and the tilde over the correlation functions indicate  Fourier transforms. The imaginary parts of this equation vanish in the Born-Markov approximation because in the master equation complex conjugate terms are summed.

For a formulation that can be implemented efficiently in source code, the quantum master equation is  rewritten in the form 
\begin{align}\frac{d}{dt} \vec{\rho}=\mathcal{M}\vec{\rho}. 
\end{align}
Here $\vec{\rho}=\text{vec}\{\rho\}$ denotes the column-vectorization of the matrix $\rho$, meaning that the $(i+1)^{\rm th}$ column of the matrix is written below the other $i$ ones. The supermatrix M has the dimension $\text{dim}(\rho)^2$, which implies an extension of the complexity, but on the other hand the solution of this equation is possible with standard numerical tools. With the help of the relation \cite{Barnett1990}
\begin{align}
  \text{vec}\{AXB\}=(A\otimes B^T)\text{vec}\{X\},
\end{align}
where $A$, $X$ and $B$ are matrices, the transformation of quantum master equation into the wanted type is possible. The symbol $\otimes$ is the Kronecker-product of the matrices is defined as
\begin{align}
 A\otimes B = \begin{bmatrix} a_{11} B & \cdots & a_{1n} B \\ \vdots & \ddots & \vdots \\ a_{m1} B & \cdots & a_{mn} B \end{bmatrix}.
\end{align}
%If $A$ and $B$ have dimension $n$, the Kronecker-product of these matrices has dimension $n^2$. 
For the problem considered in this paper we get
\begin{align}
  \mathcal{M}&=\, \mathcal{M}_C+\mathcal{M}_D\\
  \mathcal{M}_C&=\, i\left(\mathds{1}\otimes H_S^T-H_S\otimes\mathds{1}\right)\\
  \mathcal{M}_D&=\, -\sum_{\substack{\nu,\nu'=+,-,z\\ \alpha,\alpha'=T,B}}\left[\left(J_\nu S_{\nu\nu'}^{\alpha\alpha'}(+\tau)\otimes\mathds{1}\right)\right.\notag\\
  &-\left(S_{\nu\nu'}^{\alpha\alpha'\sigma}(+\tau)\otimes (J_\nu)^T\right)\notag\\
  &+\left.\left(\mathds{1}\otimes[S_{\nu\nu'}^{\alpha\alpha'}(-\tau) J_\nu]^T\right)-\left(J_\nu\otimes [S_{\nu\nu'}^{\alpha\alpha'}(-\tau)]^T\right)\right],
\end{align}
where $\mathcal{M}_C$ contains the coherent part of the master equation and $\mathcal{M}_D$ the dissipative part.

\section{Calculating the Current}
The current is given by the time-derivative of the number of particles of the tip $N_T(t)=\sum_{k\sigma}c_{k\sigma}^{T\dagger}(t)c_{k\sigma}^{T}(t)$, 
\begin{align}
 I_{Th}(t)= e\frac{d}{dt}\langle N_T(t)\rangle=ie\langle[H,N_T(t)]\rangle.
\end{align}
In the commutator $[H,N_T(t)]$ only the coupling Hamiltonian $H_C$ survives, which leads to 
\begin{align}
 I_{Th}=-ie&\sum_{kk'}t^{TB}\left(\langle J_+c_{k\downarrow}^{B\dagger}c_{k'\uparrow}^{T}\rangle+\langle J_-c_{k\uparrow}^{B\dagger}c_{k'\downarrow}^{T}\rangle\right.\notag\\
 &+\langle J_z\left[c_{k\uparrow}^{B\dagger}c_{k'\uparrow}^{T}-c_{k\downarrow}^{B\dagger}c_{k'\downarrow}^{T}\right]\rangle\notag\\
 &-\langle J_+c_{k\downarrow}^{T\dagger}c_{k'\uparrow}^{B}\rangle-\langle J_-c_{k\uparrow}^{T\dagger}c_{k'\downarrow}^{B}\rangle\notag\\
 &-\left.\langle J_z\left[c_{k\uparrow}^{T\dagger}c_{k'\uparrow}^{B}-c_{k\downarrow}^{T\dagger}c_{k'\downarrow}^{B}\right]\rangle\right) .
\end{align}
This form of the current bears similarity to terms in the master equation. It contains all the tunneling events from the tip to the bulk with a positive sign and those in opposite direction with a negative sign. Proceeding similar to the steps used for the dissipative part $\mathcal{M}_D$ of the master equation in Born-Markov approximation, and concentrating on the stationary limit one finds \cite{1407.5460}
\begin{align}
\mathcal{M}_D^I=&\sum_{\substack{\nu,\nu'}}\left\{S_{\nu\nu'}^{TB}(+\tau)\otimes J_\nu^T+J_\nu\otimes [S_{\nu\nu'}^{TB}(-\tau)]^T\right.\notag\\
 &\left.-S_{\nu\nu'}^{BT}(+\tau)\otimes J_\nu^T-J_\nu\otimes [S_{\nu\nu'}^{BT}(-\tau)]^T\right\},\\
 I_{Th}=&\text{Tr}\left[\hat{I}_{Th}\rho\right]=e\sum_{ij}(\mathcal{M}_D^I)_{(j-1)\cdot17+j,i}(\vec{\rho}_{St})_i \,.
\end{align} 
The sum over $j$ runs over all states and stands for the trace,  $1\le j\le 17$, whereas the sum over $i$ is also over all off-diagonal elements and corresponds to the matrix multiplication, thus $1\le i\le 17^2=289$. The complicated indices of $M_D^I$ in the sum are a result of the vectorization of the trace.

To estimate the 'leakage' current $I_{Leak}$ mentioned in the main part of the paper we use the fact that the leakage current does not depend on the state of the Ho atom and thus can be distinguished from $I_{Th}$. By comparing the height of the step of the differential conductance $G=dI/dU(U)$ from elastic processes at zero bias $U=0$ to voltages $U^>>\Delta E_{87}$ at energies above the first excitation gap, the leakage current can be estimated. The step size is defined as $s=G(U^>)/G(0)-1$. In the experiments a step of $s_{Exp}\approx0.9\%$ was measured. Theoretically we expect without leakage, broadening and modulation as step size of $s_{Th}\approx10\%$. But if we assume that only $10\%$ of the total current is described by $I_{Th}$, and $90\%$ is leakage current, we would expect a step size of $s\approx 1\%$, which is in good agreement. For a more quantitative analysis of the dependence on the coupling constant $c_{TB}$, and thus the current, we refer to Fig.~\ref{pLifetUMerge} c).

\end{document}